\documentclass[
    ,final            
  ]
  {aipproc}

\layoutstyle{8x11single}

\newcommand{\imedeaaddress}{Instituto Mediterr{\'a}neo de Estudios Avanzados IMEDEA (CSIC-UIB).
  \\C/ Miquel Marqu{\`e}s, 21, E-07190 Esporles, Mallorca, Spain}

\newcommand{\ufiaddress}{Unidad de F{\'\i}sica Interdisciplinar-IMEDEA (CSIC-UIB).
  \\Campus Universitat de les Illes Balears, E-07122 Palma de Mallorca, Spain}

\newcommand{\faroaddress}{CCMAR, CIMAR-Laborat{\'o}rio Associado, Universidade do Algarve\\
Gambelas, 8005-139, Faro, Portugal}

\begin{document}

\title{Evolutionary and Ecological Trees and Networks }


\author{Emilio Hern{\'a}ndez-Garc{\'\i}a}{
  address={\ufiaddress}
}

\author{E. Alejandro Herrada}{
  address={\ufiaddress}
}

\author{Alejandro F. Rozenfeld}{
  address={\imedeaaddress}
}

\author{Claudio J. Tessone}{
  address={\ufiaddress}
}

\author{V{\'\i}ctor M. Egu{\'\i}luz}{
  address={\ufiaddress}
}

\author{Carlos M. Duarte}{
  address={\imedeaaddress}
}

\author{Sophie Arnaud-Haond}{
  address={\faroaddress}
  ,altaddress={DEEP/LEP-Laboratoire Environnement Profond\\
IFREMER Centre de Brest, BP 70 29280, Plouzane,
France} 
}

\author{Ester Serr{\~a}o}{
  address={\faroaddress}
}

\begin{abstract}
Evolutionary relationships between species are usually represented
in phylogenies, i.e. evolutionary trees, which are a type of
networks. The terminal nodes of these trees represent species,
which are made of individuals and populations among which gene
flow occurs. This flow can also be represented as a network. In
this paper we briefly show some properties of these complex
networks of evolutionary and ecological relationships. First, we
characterize large scale evolutionary relationships in the Tree of
Life by a degree distribution. Second, we represent genetic
relationships between individuals of a Mediterranean marine plant,
Posidonia oceanica, in terms of a Minimum Spanning Tree. Finally,
relationships among plant shoots inside populations are
represented as networks of genetic similarity.
\end{abstract}

\maketitle


\section{Introduction}

The study of complex networks, representing interactions among
components, has become a central tool in the science of complex
systems \cite{Albert2002,Dorogovtsev2002,Boccaletti2006}.
Evolutionary relationships between species are usually represented
in phylogenies, i.e. evolutionary trees. One branching event
represents the evolution of an ancestral species into descendent
ones. The whole set of relationships among all known species is
conceptually represented as a huge {\sl Tree of Life}. A tree is a
network in which there are no cycles, i.e., there is a unique path
from one node to another inside the network. This is probably a
good approximation to the correct large scale structure of the
Tree of Life, but processes such as lateral gene transfer or
hybridization would need a richer network structure to be properly
represented \cite{Posada2001}. If analyzing the Tree of Life at a
finer detail, entering the scale appropriate for ecological
interactions, we observe that species are composed of different
populations, and that those are made of individuals that
interchange genes and recombine their genomes in processes such as
sexual reproduction. Thus, there are gene flow processes,
particularly obvious when looking at the intraspecific level,
which add loops to the Tree of Life, and make the whole structure
a rather complex object.

In this paper we briefly analyze, at three different scales, some
properties of this complex network of evolutionary and ecological
gene flow. Our aim is just to provide an overview of the many
interesting features of genetic relationships and phylogenies that
can be addressed from the point of view of complex systems, with
the hope that this will stimulate further and more detailed work.
First, we characterize large scale evolutionary relationships, the
ones more traditionally represented in a tree topology. To this
end we analyze a large scale reconstruction of the Tree of Life
and characterize it by its degree distribution. Second, we
represent the genetic relationships between individual shoots of a
Mediterranean marine plant, Posidonia oceanica, sampled across its
entire geographical extent, in terms of a Minimum Spanning Tree
that would represent the most parsimonious pattern of gene flow
between distant populations. Finally, relationships among shoots
in the same population are represented as networks of genetic
similarity.

\begin{figure}[h!t]
\includegraphics[width=.7\textwidth,angle=0]{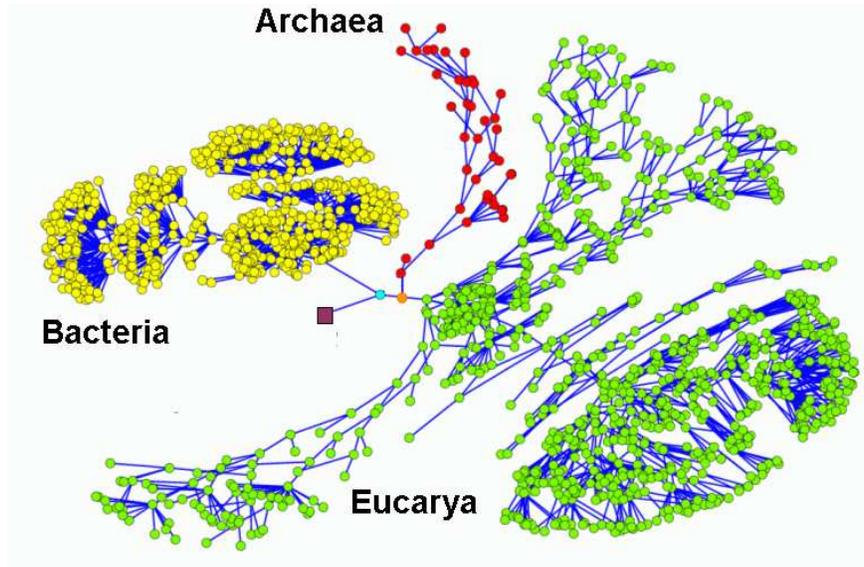}
\caption{A rendering of the first branchings of the ``Tree of Life".
Many additional subdivisions occur at each of the final branches
in this plot, where we can distinguish the three main Domains
(Bacteria, Archaea, Eucarya) in which the different organisms are
classified. }
\label{fig:Tree of Life}
\end{figure}


\section{Scale-free degree distribution in the Tree of Life}
\label{sec:Tol}

Since the beginning of the studies on evolutionary biology, one of
the most ambitious goals has been the assemble and comprehension
of the complete evolutionary history of biodiversity. This
assembly has produced a huge phylogenetic tree, baptized with the
name ``Tree of Life". The data set analyzed here is the
reconstruction of the Tree of Life which is available at the
database of the {\sl Tree of Life Web Project}
(\url{http://www.tolweb.org/}). It contains about $6 \times10^5$
nodes. A very small portion of it, showing the first
ramifications, is shown in Fig.~\ref{fig:Tree of Life}.

\begin{figure}[h!b]
\includegraphics[width=.5\textwidth,angle=-90]{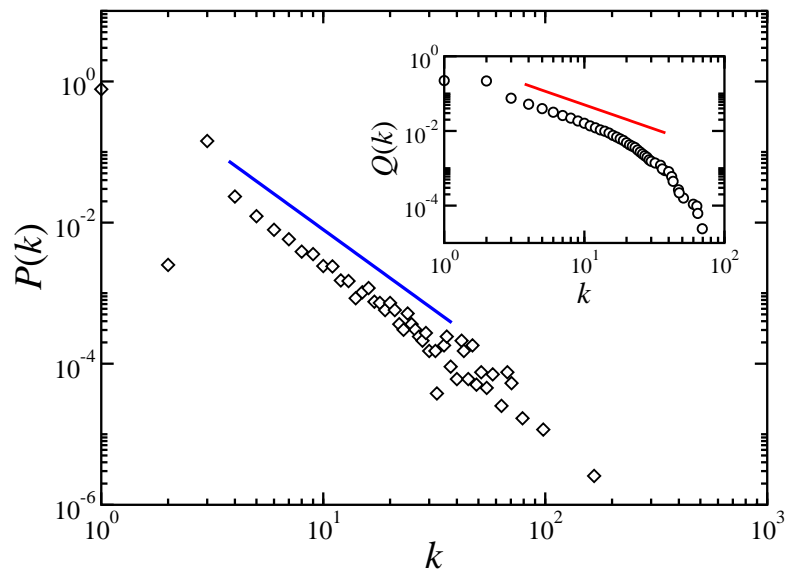}
\caption{Degree distribution in the Tree of Life. The fit is a power law
$P(k)\sim k^{-\gamma}$ with the scaling exponent $\gamma \approx
2.33$. The inset shows it in cumulative form: $Q(k)=1-\sum_{r=1}^k
P(r)$. }
\label{fig:Degree distribution}
\end{figure}


One of the first quantities to discuss when analyzing complex
networks is the degree distribution, $P(k)$, i.e.  the proportion
of nodes in the network which are connected to $k$ neighbors. For
our tree, this function is plotted in Fig.~\ref{fig:Degree
distribution} (together with the function $Q(k)=1-\sum_{r=1}^k
P(r)$, giving an accumulated version of the distribution). We see
that the probability of finding a node with degree $k$ decays as a
power law $P(k) \sim k^{-\gamma}$, with $\gamma \approx 2.33$.
Thus, this phylogenetic tree has a broad degree distribution of
the scale-free type. The scaling exponent $\gamma$ characterizing
this distribution is consistent with the results obtained by
Cartozo {\sl et al.} \cite{Cartozo2006} from a previous taxonomic
diversity analysis, where the values obtained were in the range
$1.9-2.7$.

The use of the degree distribution to characterize a phylogenetic
tree needs some qualification:  It is generally accepted that a
true phylogenetic tree at high enough resolution will contain only
binary branchings \cite{Mooers1997}. This corresponds to $k=3$ for
all internal nodes, and $k=1$ for the root and the terminal tips,
that would constitute about one half of the nodes. The presence of
nodes with a very high degree (see Fig.~\ref{fig:Degree
distribution}) in our data set reveals a large amount of
polytomies. This would correspond to successive branchings of
several species or groups in an order that can not be resolved
with the techniques used in the phylogenetic reconstruction, so
that all of them are assigned the same branching point. In this
sense, it can be thought that a characteristic such as the degree
distribution is quantifying more the limitations of the
methodology used to reconstruct the phylogenetic history than the
true evolutionary branching itself. The limitation would be
clearer if considering the statistics of taxonomic
classifications, in which branchings are forced to fit into the
restricted set of levels established by taxonomic science (Domain,
Kingdom, Phyllum, Class, Order, Family, Genera, Specie). The {\sl
Tree of Life Web Project} used here does not use a strict
taxonomic classification, since new levels are added as needed. It
is likely that the broad distribution of degrees obtained here
reflects a property of {\sl radiation} processes during
evolutionary history, i.e. events in which many new species
appeared during a relatively short period of time
\cite{Givnish1997}. Clearly, further analysis is needed to
establish which features of the degree distribution arise from
phenomena of biological relevance and which are consequences of
the lack of resolution of available reconstruction methods. In any
case, it is seen in Fig. \ref{fig:Degree distribution} that the
proportion of nodes with $k=3$, representing binary branchings,
stands up above the background power law behavior. We see also the
presence of some nodes with $k=2$ (they represent taxa which
consist of only one subtaxon), although in a very small
proportion.

\section{Inside one leaf of the tree: genetic relationships in a plant species}

We now consider in more detail the genetic structures present at
one of the tips of the above Tree of Life. We focus in a
particular species and analyze its internal genetic relationships,
first across its whole geographical range, and then at particular
locations.

\begin{figure}[ht]
  \includegraphics[width=.6\textwidth]{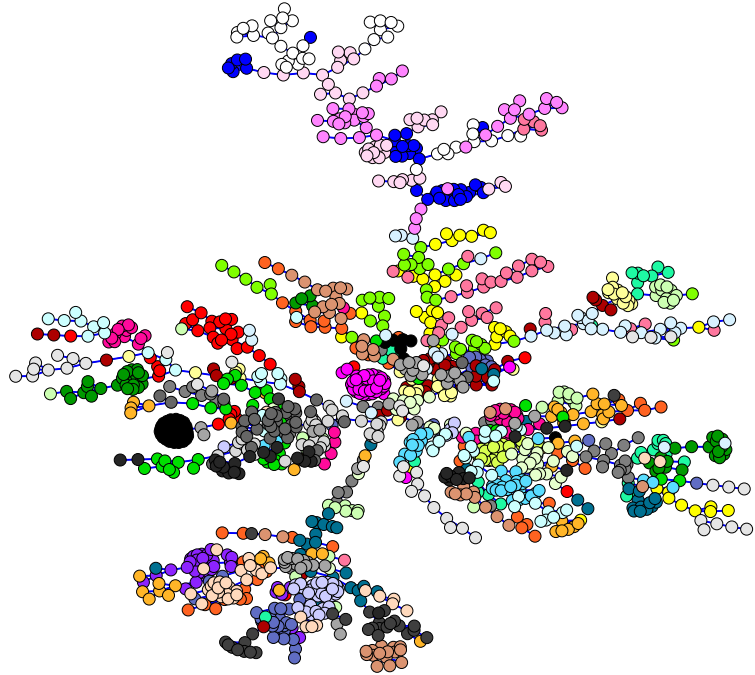}
  \caption{Minimum spanning tree containing all
sampled ramets of Posidonia oceanica. Each node represents a
shoot, with colors (or grey shades) indicating the sampling
meadow, and each link has an associated distance. The minimum
spanning tree minimizes the total sum of distances.}
\label{fig:MST}
\end{figure}

We consider the case of Posidonia oceanica, a marine angiosperm
living in meadows submerged between 0 to 40 m in the Mediterranean
Sea \cite{Hemminga2000}. It combines clonal reproduction with
episodes of sexual reproduction. This plant is experiencing
basin-wide decline and is subject to specific protection and
conservation measures \cite{Hemminga2000}. Genomic DNA data are
available from approximately 40 shoots sampled in each of 37
localities across the Mediterranean. A set of seven microsatellite
markers was used to characterize the genotype of each individual.
This provides us with a data set
\cite{Arnaud-Haond2005,Rozenfeld2006,HernandezGarcia2006}
consisting on the number of repetitions of the microsatellite
motif at each locus of each shoot sampled. In this set, a
convenient genetic distance $d_{ij}$ can be defined
\cite{Rozenfeld2006,HernandezGarcia2006} which measures the degree
of dissimilarity among every pair $\{i,j\}$ of sampled shoots.
From this distance matrix, several trees and networks can be built
and analyzed, as described in the following.

\begin{figure}[h!b]
  \includegraphics[width=.6\textwidth,angle=-90]{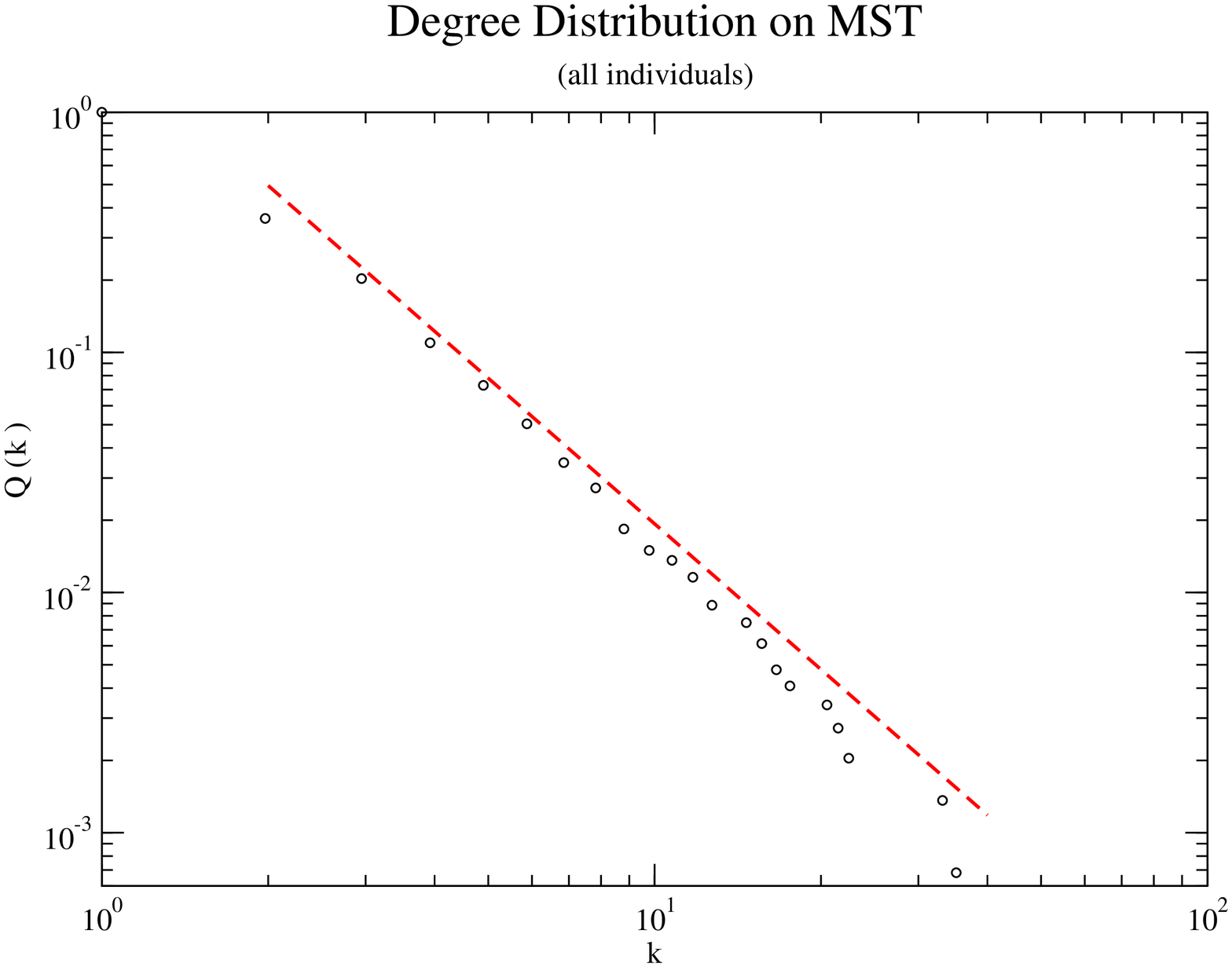}
  \caption{Representation of the cumulative degree distribution
$Q(k)=1-\sum_{r=1}^k P(r)$ characterizing the MST. The straight
line is a fit to $Q(k)\sim k^{-\gamma+1}$, with $\gamma\approx
2.95$}
  \label{fig:degreeMST}
\end{figure}

\subsection{The minimum spanning tree of plant shoots}
\label{subsec:MST}

We can consider the fully connected network linking all pairs of
shoots in our data set. Given a connected, undirected graph, a
spanning tree of that graph is a subgraph which is a tree, i.e.
contains no loops, and connects all the vertices together. A
single graph can have many different spanning trees. Our distance
matrix $d_{ij}$ assigns a weight to each of the links. We can use
this information to assign a total {\sl length} to each of the
possible spanning trees by computing the sum of the distances of
the edges which are kept in that spanning tree. A minimum spanning
tree (MST) is then a spanning tree with the minimum possible total
length. In the case in which all edges have a different distance,
the MST is unique. But more generally, there could be several MSTs
for a particular network and distance matrix. A MST is in fact the
minimum-cost subgraph connecting all vertices, since subgraphs
containing cycles necessarily have more total weight. This allows
to interpret the MST as the main path of gene flow among the plant
populations, as it links all the specimens in the genetically
shorter way.

Fig. \ref{fig:MST} shows the MST associated to the set of shoots
and the corresponding distances (it is in fact one of the several
equivalent MST with identical total length and similar shape that
can be constructed from our distance matrix). The nodes,
representing shoots, are colored with the same shade if they have
been collected from the same meadow. The upper part of the tree
contains shoots from populations in the Eastern and Central
Mediterranean, and the lower part from the Northern Spanish coast,
in the Western Mediterranean. The central part of the MST is
occupied by populations in the Balearic Islands. The tightly
packed balls of nodes group together genetically identical shoots,
arising from clonal reproduction. Although most of the shoots from
the same location appear close together in the tree, there is some
scatter, indicating that there has been some migration between
different populations. This would need to be represented with
additional links introducing cycles and transforming the tree into
a more complex network.

Figure \ref{fig:degreeMST} shows the cumulative degree
distribution characterizing the MST. It follows a power law with
exponent -1.95, so that the degree distribution is of the
scale-free type, satisfying $P(k) \sim k^{-2.95}$.

\subsection{Intrapopulation networks}
\label{subsec:PopNets}

\begin{figure}
  \includegraphics[width=.8\textwidth]{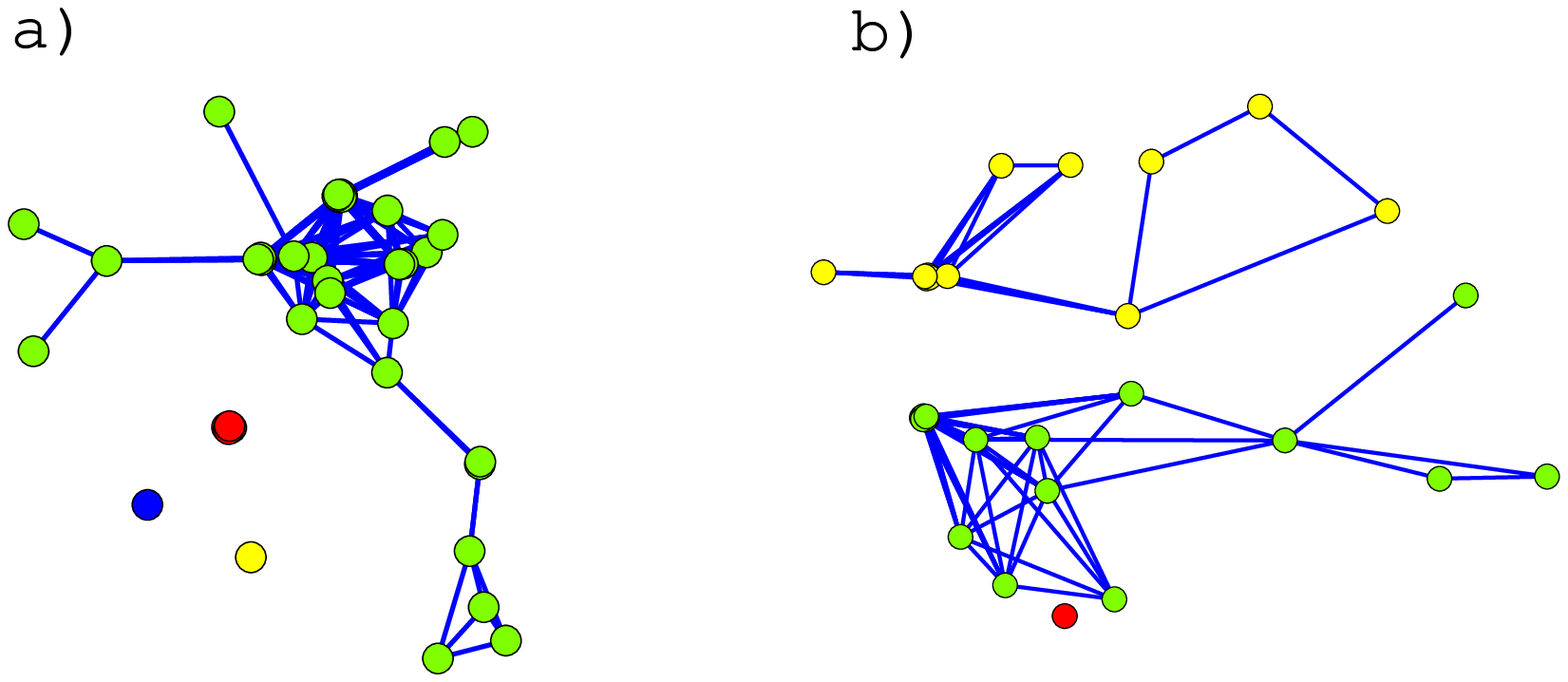}
  \caption{Networks of genetic similarity constructed for the shoots sampled
at a) Es Pujols (Formentera, Balearic Islands), and b) Campomanes
(Mediterranean Spanish coast). }
\label{fig:EsPujolsCampomanes}
\end{figure}

Still at a smaller scale, we can address the genetic relationship
among shoots sampled inside the same meadow. It is clear that a
tree structure is not appropriate to represent the intense gene
mixing that would be provided by sexual reproduction at this
scale. Starting from the fully connected network containing links
among all pairs of shoots, we can discard the links containing
higher distances and retain only those that represent a
sufficiently close genetic similarity. This is related to the
network construction methods based in correlation thresholds
\cite{Onnela2003,Eguiluz2005}. In \cite{Rozenfeld2006} we
constructed networks of genetic similarity among {\sl genets} (the
set of clonally identical shoots) by choosing as the distance
threshold the so called {\sl outcrossing distance}. This is the
average distance between the genets in a population and its
offspring obtained from a simulation of outcrossing, i.e. sexual
reproduction among genetically different individuals. The same
idea is applied here to obtain networks of shoots. Network links
will join shoots which are genetically closer than the typical
outcrossing distance for their population. Figure
\ref{fig:EsPujolsCampomanes} shows two examples of networks
obtained in such a way. The network representation visually
highlights the main features of the population structure. For
example the population in Es Pujols (Fig.
\ref{fig:EsPujolsCampomanes}a) consists of a central core of
interconnected shoots to which less central organisms are linked.
Campomanes (Fig. \ref{fig:EsPujolsCampomanes}b), instead, is
structured in two main components. The size of these networks is
too small to search for scaling properties in the degree
distribution, but the analysis in \cite{Rozenfeld2006} at the
genet level reveals that they have the characteristics of small
worlds \cite{watts1998}, i.e. clustering significatively higher
than a random network with the same number of nodes and links,
while at the same time keeping the same low diameter values
characteristic of random networks.

\section{Conclusion}

We have presented a brief overview of properties of trees and
networks constructed in the context of phylogenetic and genetic
relationship at very different scales, from evolutionary to
ecological. The visual inspection of these structures reveals
interesting clues such as paths of gene flow, patterns of
speciation, or population structure. The open challenge is to
relate more quantitative topological properties of these complex
networks to relevant biological mechanisms.


\begin{theacknowledgments}
We acknowledge financial support from the Spanish MEC (Spain) and
FEDER through projects CONOCE2 (FIS2004-00953) and SICOFIB
(FIS2006-09966), the Portuguese FCT through project NETWORK
(POCI/MAR/57342/2004), the BBVA Foundation (Spain), and the
European Commission through the NEST-Complexity project EDEN
(043251).
\end{theacknowledgments}






\end{document}